\documentclass[%
 reprint,
 amsmath,amssymb,
 aps,
]{revtex4-1}

\usepackage{graphicx}% Include figure files
\usepackage{dcolumn}% Align table columns on decimal point
\usepackage{bm}% bold math
\usepackage{hyperref}% add hypertext capabilities
\usepackage[english]{babel}

\newcommand{\ld}{\lambda_\mathrm{D}}
\newcommand{\ls}{\lambda_\mathrm{S}}
\newcommand{\lb}{\lambda_\mathrm{B}}
\newcommand{\kbt}{k_\mathrm{B}T}
\newcommand{\um}{\mu_\mathrm{m}}
\newcommand{\tc}{\tau_c}

\newcommand{\comment}[1]{}

\begin{document}

%?
%\preprint{APS/123-QED}

\title{Electric-Double-Layer-Modulation Microscopy} %v2edit

\author{Kevin Namink,$^{1}$ 
Xuanhui Meng,$^{2}$ 
Marc T. M. Koper,$^{3}$ 
Philipp Kukura,$^{2}$ 
Sanli Faez$^{1,}$}
\altaffiliation{s.faez@uu.nl}
\affiliation{$^{1}$Nanophotonics, Debye Institute for Nanomaterials Research, Utrecht University, NL\\
$^{2}$Department of Chemistry, Physical and Theoretical Chemistry Laboratory, Oxford University, UK\\
$^{3}$Leiden Institute of Chemistry, Leiden University, NL}

%\collaboration{ }%\noaffiliation

\date{\today}

\begin{abstract}
The electric double layer (EDL) formed around charged nanostructures at the liquid-solid interface determines their electrochemical activity and influences their electrical and optical polarizability. 
We experimentally demonstrate that restructuring of the EDL at the nanoscale can be detected by dark-field scattering microscopy. 
Temporal and spatial characterization of the scattering signal demonstrates that the potentiodynamic optical contrast is proportional to the accumulated charge of polarisable ions at the interface and its time derivative represents the nanoscale ionic current. 
The material-specificity of the EDL formation is used in our work as a label-free contrast mechanism to image nanostructures and perform spatially-resolved cyclic voltametry on ion current density of a few attoamperes, corresponding to the exchange of only a few hundred ions. 
\end{abstract}

%? PACS will do later

\maketitle
%\tableofcontents

\section{\label{sec:introduction} INTRODUCTION}

The storage and recovery of energy in batteries, solvation of molecules, filtration by membranes, and many transport processes in liquid environments are dictated by the interaction of ions with charged surfaces and the formation of the electric double layer (EDL)\cite{lyklema_book_1995}. 
The EDL consists of a layer of ions in solution that screens the surface charge at the interface, with a thickness from less than one to a few tens of nanometers dependent on the ionic strength of the solution. 
The formation of the EDL involves several time scales\cite{bazant_dynamics_2004} such as the relaxation time $\tau_D = \ld^2/D$, with $D$ the diffusion constant and $\ld$ the Debye length, and the charging time $\tau_c = \ld L/D$, with $L$ the representative system size. %v3edit
The Debye length is often used as a measure of the EDL thickness. 
The small volumes and short timescales associated with the formation of the EDL makes direct access to its local dynamics experimentally challenging. Previous experimental observations of the EDL on the nanoscale have been based on amperometric measurements with scanning probe methods\cite{collins_screening-dynamics_2014, bentley_echem-mapping_2018}, nanopores\cite{smeets_salt_2006, haywood_fundamental_2015}, or ultramicroelectrodes\cite{xiao_microelectrode_2007, albrecht_electrochemical_2018}, which require a current signal above the background thermal current fluctuations. 
Visualising the contrast of the EDL optically, on the other hand, probes the accumulated charge and provides direct access to spatial information of the ionic current. 
Optically studying the spatial ion accumulation and transport is an enabling approach, built on a distinct working principle, which combines the power of optical microscopy with electrochemical amperometric analysis. 

Changes to the optical reflectivity of a homogeneous flat electrode in contact with an electrolyte as a function of its electric potential, referred to as electroreflectance, has been observed previously, and attributed to modulations of the optical properties of both the metal and electrolyte\cite{hansen_electromodulation_1968, bewick_studies_1973}. 
The connection between electroreflectance and the restructuring of the EDL has been postulated and tested for large flat surfaces using ellipsometry\cite{stedman_effect_1968} but separating the changes caused by conduction electrons in the metallic layer to electromodulation from the contribution of electrolyte EDL has been challenging\cite{mcintyre_electrochemical_1973, kolb_observation_1981}.
More recently, the influence of electrode potential on elastic light scattering (ELS) from plasmonic nanoparticles that exhibit a localised plasmon resonance\cite{atwater_tuning_2015, hill_singleAg_2015, landes_single-particle_2016} and for two-dimensional materials\cite{tao_mos2contrast_2019} has been detected. 
In those experiments, the signal from the EDL is not separated from the plasmonic and electronic effects caused by variation of the charge density inside the nano-object or at the surface. 
The influence of the EDL on the scattering cross section of 5-nm silica nanoparticles have been indirectly measured by changing the salt-concentration in a quantum-noise limited measurement\cite{mauranyapin_evanescent_2017}. %v2edit

Here, we experimentally demonstrate that continuous tuning of the the EDL composition can be directly visualised by measuring the ELS from any type of nanoparticle and even surface roughness on top of a capacitively charged surface. %v2edit
We refer to this intensity change in the ELS as the potentiodynamic optical contrast (PDOC).
The temporal response of the PDOC is influenced mostly by the physical adsorption of counter-ions with different optical polarizability (related to bulk refractive index) compared to the solvent. 
We demonstrate this effect by quantifying the temporal relaxation of the PDOC, which is directly related to the charging time of the EDL.
We also show that the magnitude of PDOC is related to optical polarizability of the ions.
%At cell potentials exceeding roughly 1~V, other surface reactions influence the PDOC and its behavior no more only dependent on the EDL reconfiguratioin. 
We observe that deposited nanoparticles from other materials exhibit a different pattern than the underlying ITO substrate due to different electrochemical properties. 
This difference enables visualisation of small nanoparticles that otherwise cannot be differentiated from background scattering. 
By accurately measuring the PDOC as a function of applied potential, we can perform the optical equivalent of cyclic voltametry, at attoampere current level, on a single nanoparticle. 

\section{\label{sec:pdoc} POTENTIODYNAMIC OPTICAL CONTRAST OF THE EDL}

In this section, we present an estimation for the expected PDOC that is caused by the change in the ion concentrations inside the EDL as a function of the surface potential. 
Details of this derivation are presented in Appendix~\ref{app:pdoc}. 
We use a nanosphere with a uniformly charged surface as a model system. 
Because we are mainly interested in the ELS from the EDL, we only consider changes due to the reconfiguration of ions outside the particle. 
The optical contrast of the EDL can be used to study dielectric particles, semiconductor nanocrystals, and metallic scatterers with plasmon resonance frequencies far from the visible range.

Because the EDL is much thinner than the wavelength of the incident light, the details of the charge distribution inside the EDL has a negligible influence on the ELS intensity. 
This model matches the physical conditions for surface potentials much larger than the characteristic potential $\kbt/e \approx 25~$mV, in which charge screening is mostly due to the Stern layer. 
The total number of excess counter-ions, $N$, necessary for screening the nanosphere at surface potential $V$ is given by $(\frac{V e }{\kbt})(\frac{a^2}{\lb\ls})$, with $\lb$ the Bjerrum length and $\ls$ representing the thickness of the charge screening layer. %v2edit
In the Rayleigh scattering regime, the polarizability of the combined system of the nanosphere and the EDL is a volumetric sum of its constituents. 
Using the Rayleigh polarizability and the phenomenological linear relation between refractive index and salt concentration $n_\mathrm{mix} = n_w + K x_s$, with $x_s$ the ratio between number density of salt ions and solvent molecules\cite{an_combined_2015} and $n_w$ the refractive index of water, we arrive at the scaling result, 
\begin{equation}\label{eq:aEDL}
\frac{\alpha_\mathrm{EDL}}{\alpha_p} \simeq \frac{2 K (m^2+2)}{n_w(m^2-1)}.\frac{\frac{V e}{\kbt}}{4 \pi a \lb\ls \rho_w},  %v2edit
\end{equation}
with $\rho_w$ the number density of water molecules (considering an aqueous solution) and $m$ the ratio between refractive index of the particle and water. 
For a typical dielectric material ($m=1.3$) and alkali-halide salts\cite{an_combined_2015}, the prefactor is roughly 3. 
Using $\rho_w = 33\,\mathrm{nm}^{-3}$ and $\lb = 0.7~$nm for water and a typical $\ls = 1~$nm we arrive at $\alpha_\mathrm{EDL}/\alpha_\mathrm{p} \approx 0.04$ for a 10~nm (radius) nanoparticle in a NaCl solution at a surface potential of $V = 1$~Volt. %v2edit

\section{\label{sec:dfsm} EDL-MODULATION MICROSCOPY}

To image such small changes in polarizability due to the reconfiguration of the EDL, we use a customized total-internal reflection optical microscope. 
Similar to other interferometric-enhanced imaging techniques\cite{kukura_high-speed_2009, arroyo_non-fluorescent_2016}, the static scattering from the nanoparticle acts as a reference for the homodyne detection of the changes in the polarizability of the (sub-wavelength) particle surrounding. 
Ultimately, the imaging resolution is fixed by the optical diffraction. 
Therefore, the measured signal in each difraction-limited spot is proportional to the total change of the EDL polarizability in the diffraction volume.  %v2edit
It is essential that the reference intensity is kept stable with fluctuations smaller than the scattering contribution from the EDL. For our measurements the signal to reference ratio is on the order of $10^{-3}$ to $10^{-2}$.

\begin{figure}[ht]
\centering
\includegraphics[width=0.5\textwidth]{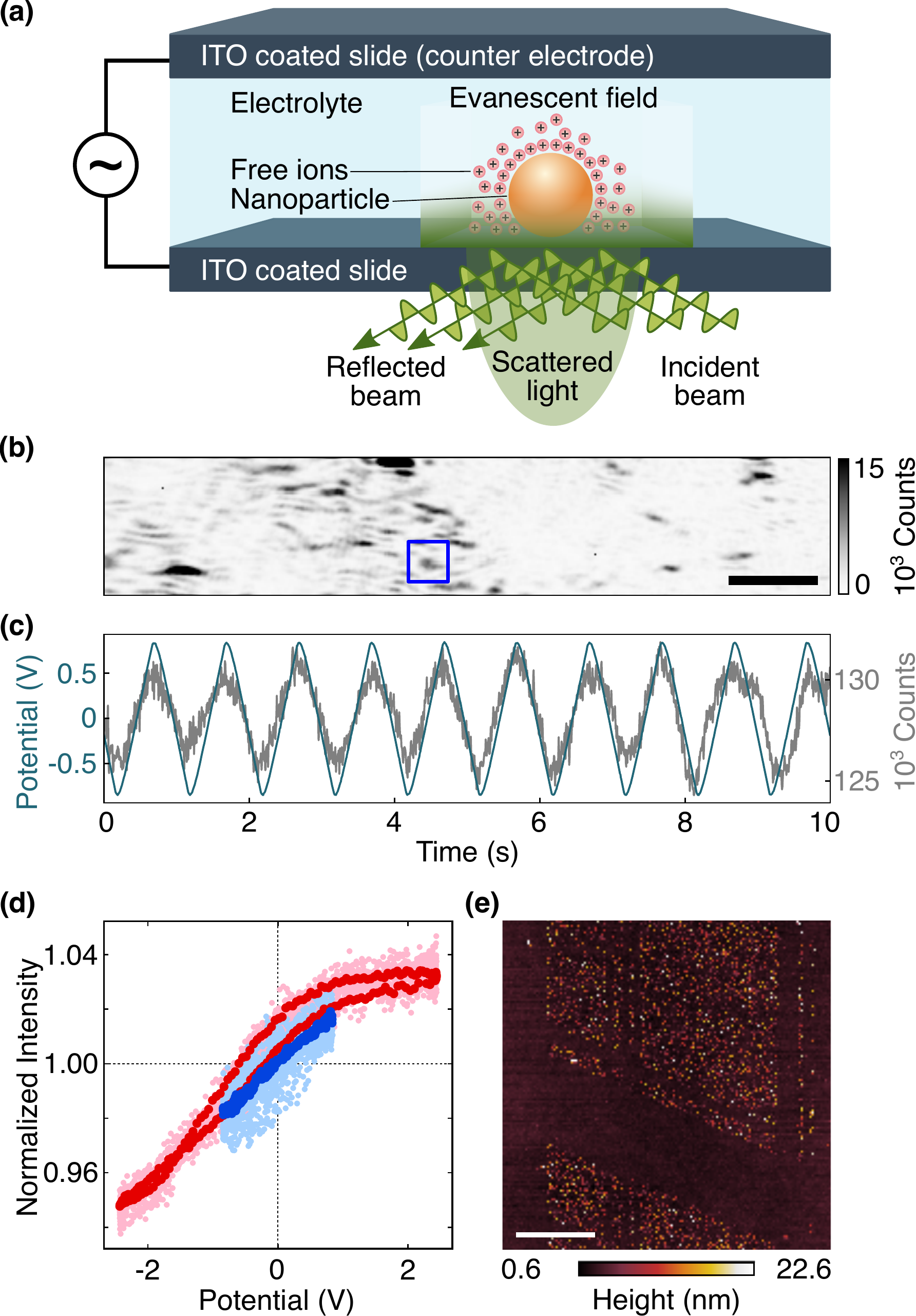}
\caption{\label{fig1setup}
Measuring the potentiodynamic contrast of the EDL on a rough surface 
(a) Setup for measuring the PDOC and the electrochemical cell configuration, 
(b) Typical scattering image from the ITO surface (scale bar: $2\um$)
(c) Scattering intensity from the grain annotated by an square in (b) plotted while changing the surface potential in a triangle waveform. 
(d) The normalized intensity change plotted as a function of the cell potential for 100 cycles (light symbols) and the average of all cycles after correction for drift (bold symbols) for two different sweeping amplitudes.
(e) Atomic force microscope scan of the ITO surface (scale bar: $4\um$) 
}
\end{figure}

\section{\label{sec:results} RESULTS}

We perform PDOC measurements on nanoparticles or grown nanostructures on top of ITO-coated glass cover-slips (Diamond Coatings, 70 - 100 Ohms/Sq.). 
As the counter-electrode we use a second ITO-coated acrylic sheet separated from the substrate using double-sided adhesive tape, forming a 100-$\mu$m-thick flow cell. 
This configuration enables liquid exchange inside the flow-cell while investigating the same field of view of the substrate with different electrolyte solutions. 
Unless specified otherwise, the applied potentials have balanced triangle-shape waveform and the scattering images are recorded at 200 frames per second. 

In Fig. \ref{fig1setup}, we present a typical scattering image of the surface of the ITO substrate. 
The ITO surface contains regions of high scattering intensity in the shape of parallelograms (see Fig. \ref{fig1setup}(b)), with sharp edges and corners, in the middle of comparatively smoother regions of 10 to 100 times lower scattering signal. 
Using an atomic force microscope, we could detect the presence of sparse grains of roughly 20~nm high in geometrically recognisable areas (Fig.~\ref{fig1setup}(e)). 
The sharp boundaries confining these grains are indicative of the crystallographic origin for their formation, attributed to the stress release in the deposited layer during the annealing of ITO\cite{kavei_evaluation_2008, gardonio_characterization_2008}. 
We can use the fairly homogeneous size distribution of these grains on the ITO rough regions, and straight boundaries of these regions to distinguish between ITO grains and other particles or contamination that resides on the surface. 
For the cell potentials and electrolyte solutions used here, the ITO surface has proven to be very stable and has shown no irreversible change for phosphate buffer at pH 7, when the cell potential is kept within $\pm 1.5$ V. 

Due to their electrochemical stability, we can use the ITO nanograins as reference scatterers for measuring the EDL signal.
We record the ELS while alternating the potential of the ITO substrate relative to the counter electrode. 
By subtracting the average scattering signal over an entire cycle from each frame we obtain the PDOC and simultaneously correct for any drift in background intensity. 
For some of the spots, these intensity oscillations are visible above the measurement noise, even for a single cycle, after correcting for the drift. 
Fig.~\ref{fig1setup}(b) depicts the average dark-field scattering image of the ITO surface.
The ELS intensity from a single speckle spot is plotted in Fig.~\ref{fig1setup}(c) as a function of the applied potential for ten cycles. 
For low cell potentials, the relation between contrast and substrate potential is close to linear at any instant, in agreement with the prediction for EDL restructuring. 
At higher potentials, however, we observe a nonlinear dependence and a phase lag between the PDOC and the applied potential (Fig. \ref{fig1setup}(d)).

\begin{figure}[ht]
\centering
\includegraphics[width=0.5\textwidth]{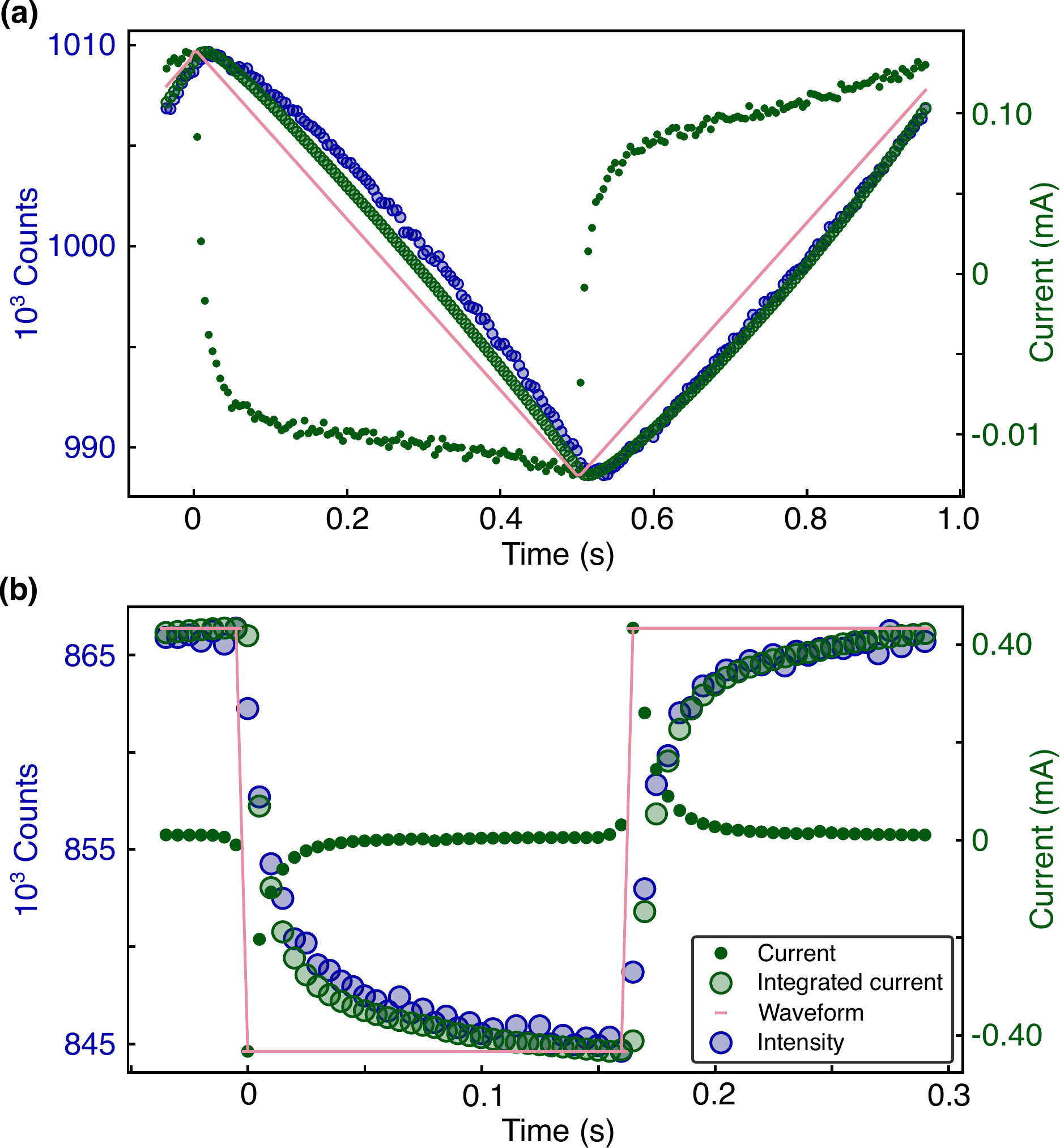}
\caption{\label{fig2response}
The optical signal from the electric double layer under variable surface potential: The measured electric current through the cell for an alternating voltage of (a) triangle and (b) square waveform between +1 and -1 V. 
In both (a) and (b) in green dots the measured current and in pink dots the applied potential. The integrated current (green circles), corresponding to the accumulate charge at the interface shows the same temporal behaviour as measured scattering intensity (blue circles, averaged value over several cycles).
The exponential change of the current towards equilibrium corresponds to the charging time of the electrochemical cell. 
}
\end{figure}

Next, we investigate the time dependence of the PDOC for a linear sweep and for a step reversal of the substrate potential, while simultaneously measuring the current passing through the cell. 
For a linear sweep, the electric current reaches a constant value after a certain relaxation time.
The PDOC follows the potential with a lag that is comparable to that relaxation time and matches the charging time $\tc$ of the flow-cell. 
This relaxation behaviour is more evident when applying a square potential, in which the current stops after $\tc$, due to screening and the PDOC saturates. 
The saturation of the current for a linear potential sweep and relaxation to null for a step change in the potential both point towards the absence of sustained Faradaic currents at the electrodes.
To make a direct comparison, we plot the integrated electric current during the cycle, i.e. the accumulated charge, on top of the PDOC signal (Fig.~\ref{fig2response}).
We observe an almost perfect overlap that the PDOC is proportional to the accumulated charge at the substrate.

\begin{figure}[ht]
\centering
\includegraphics[width=0.5\textwidth]{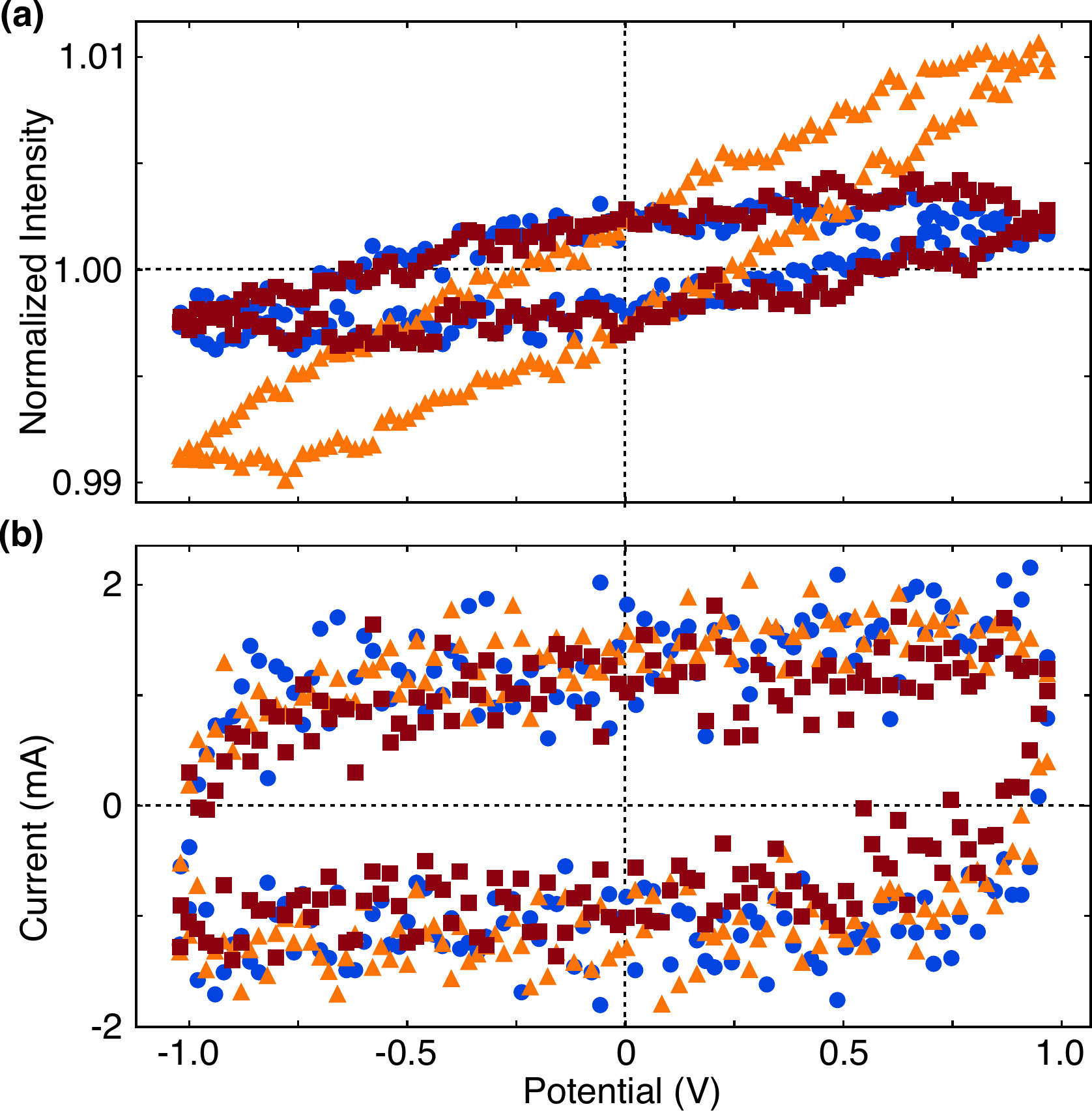}
\caption{\label{fig3_salts} 
(a) The average PDOC contrast for the same ITO grain for three different salts and
(b) the electric cyclic voltagram of the ITO substrate of the same measurements as (a).
The measured salts are NaCl (blue circles), NaBr (red squares), NaI (orange triangles)}
\end{figure}

While the above observations demonstrate the surface charging origin of the PDOC, they are insufficient to distinguish between reconfiguration of the EDL and the redox reactions at the surface of the ITO, also known as pseudocapacitance charging \cite{bard_electrochemical_2008}. 
To differentiate between these two effects, we investigate the PDOC response on the same ITO grain for three different anions in the electrolyte solution. 
Typical responses are shown in Fig.~\ref{fig3_salts}(a), next to the simultaneously measured electric current passing through the cell. 
While the electric current is the same for the three ions, the PDOC in presence of NaI is almost twice that of NaCl and NaBr for the same cell potential.
This observation can be explained by the optical polarizability of the iodide ions relative to chloride and bromide. 
While the exact calculation of the change in the refractive index of the EDL would require an accurate consideration of the ion hydration and is beyond the scope of this article, it has been shown empirically that the change in the refractive index is almost proportional to the atomic polarizability\cite{an_combined_2015}. 

%NaCl($\alpha_{Cl}$ = 14.6, RI = 1.54), NaBr ($\alpha_{Br}$ = 21, , RI = 1.64), NaI ($\alpha_{I}$ = 32, , RI = 1.78)\cite{rii}.

We also observe a difference in the temporal phase lag between in the PDOC response and the cell potential for the three electrolytes.
This difference is also observed in the electrical measurements and can be attributed to the difference in surface adsorption dynamics for the three ions. The I-V curves measured simultaneously also exhibit this difference. 

We have shown, hitherto, that the PDOC obtained by dark-field ELS microscopy is an optical indicator of the optical polarizability of accumulated ions (charges) around ridges or grains on a flat substrate. 
Furthermore, we observe that both the temporal hysteresis behavior of the cyclic optical contrast depends on the type of salt and the sweeping rate of the cell potential. 
As such, the PDOC of each grain can be viewed as a local nano-electroscope placed directly on the surface that can be used for studying the heterogeneity of surface interactions with the electrolyte, akin to conventional cyclic voltametry. 
While for electrochemically inert ITO, the EDL reconfiguration is the main source for changes to the scattering intensity, other material-specific surface reactions can influence the dynamics. 
As such, the PDOC can be seen as a novel, material-specific contrast mechanism.
To showcase this specificity, we perform EDL-modulation microscopy of chromium nanoparticles deposited on ITO.  %v3edit

\begin{figure}[ht]
\centering
\includegraphics[width=0.5\textwidth]{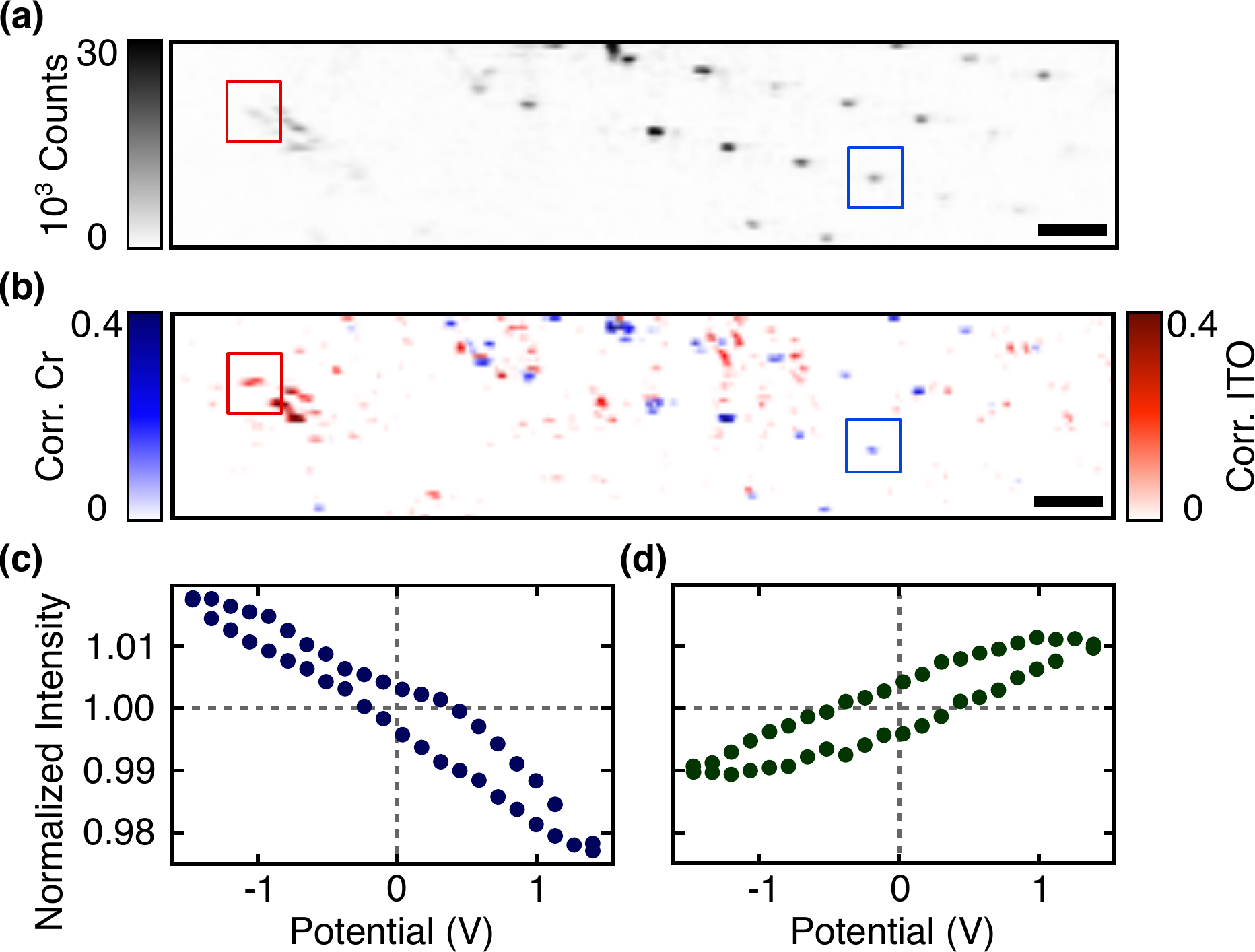} %v2edit
\caption{\label{fig4grid} 
(a) Dark-field scattering image of the ITO substrate after deposition of Chromium through a SiN stencil (scale bar: $4\um$). The deposition locations are on a triangular lattice. 
(b) Average covariance of each pixel intensity with the PDOC curve of a chromium particle is blue colorscale and covariance with the PDOC curve and ITO grain is depicted in red. 
The corresponding reference PDOC curves are depicted in panels (c) and (d). 
All pixels with a covariance of less than 0.1 with either of the two references are colored in white for clarity.  %v2edit
}
\end{figure}

To fabricate a recognizable pattern, we deposit a few nanometers of chromium on the ITO-coated slides through a SiN membrane containing an array of micrometer-size holes, used as a stencil. 
In Fig.~\ref{fig4grid}(a) we depict the recorded scattering image. 
The ELS from chromium deposits is comparable in magnitude to the ITO grains in the rough regions. The chromium deposits can be identified from their geometrical arrangement on a triangular lattice, dictated by the stencil. 
The average PDOC over several cycles for one of chromium particles and one ITO grain are depicted in Fig.~\ref{fig4grid}(c,d). 
We attribute this difference to electro-oxidation of chromium deposits. 
We can identify all other positions on the surface that exhibit the same PDOC response by correlating each pixel intensity over time with the obtained reference. 
In Fig.~\ref{fig4grid}(b) we depict the covariance of each pixel with the two different references for corresponding to Cr and ITO particles, in blue and red correspondingly. 
The position of the Cr deposits, colored in blue, matches the pattern expected from triangular mesh used for deposition. This separation between the two materials could not be done based on the scattering signal in panel (a). %v2edit
%The PDOC curve of each individual spot is depicted in the supplementary information.

\section{\label{sec:discussion} DISCUSSION}

To conclude, we have spatially resolved the reconfiguration of EDL directly from changes to the optical contrast. 
At low potentials compared to the electrochemical reaction potential, using fully polarizable electrodes, the potentiodynamic scattering contrast is mostly due to the reconfiguration of the EDL.
For higher potentials, surface adsorption or Faradaic reactions start to dominate changes in the optical contrast of the surface surrounding. 
EDL-modulation microscopy can thus be used to measure the deposition or formation of products on the surface. 
In this range, the shape and magnitude of the response depend on the sweeping rate. 
Given that particles as small as 10~nm have been detected, the local current passing through this area for the slowest scan rates in our measurements is at the level of $10^{-18}$ A. 
By further studying the correspondence between different sweeping modes, EDL-modulation microscopy can build upon the vast knowledge obtained from electrochemical studies. 
This imaging technique provides important additional information such as spatial resolution, sensitivity to surface heterogeneity, local ion accumulation, and possibility of studying deposits, possibly down to single biomolecules. 
Another operation mode compatible with EDL-modulation microscopy involves a substrate covered with a thin insulating layer. 
In this mode, the ion configuration at the EDL can be altered by capacitive coupling, and the Faradaic reaction will be completely excluded. 
This possibility will extend applications to a range of electrolytes that are chemically corrosive for ITO.

While we have chosen dark-field imaging for this work, using the EDL-modulation as an optical contrast is fully compatible with interferometric scattering microscopy (ISCAT)\cite{kukura_high-speed_2009, young_kukura_annurev-physchem}.
This method has already proven to be sufficiently sensitive for detecting and characterizing single proteins based on their polarizability\cite{young_iscams_2018}.
The combination of this remarkable sensitivity with potentiodynamic control creates a previously untapped contrast mechanism for chemical specific optical microscopy of single nanoparticles and single macromolecules. 
This possibility paves the way to measuring chemical reactions such as oxidation and reduction processes on a single protein, or their reaction with antibodies, for an extended period of time.

\begin{acknowledgments}
We thank Christian Post for performing the atomic force microscopy scans of the ITO surface and Allard P. Mosk, Serge Lemay, and Willem Boon for fruitful discussions.
This research was supported by the Netherlands Organization for Scientific Research (NWO grant 680.91.16.03). PK was supported by an ERC Consolidator Grant (819593).
These results are subject to a priority patent application, UK 1903891.8, filed on 21st March 2019.

%Here you should list the contents of your Supplementary Materials -- below is an example. 
%You should include a list of Supplementary figures, Tables, and any references that appear only in the SM. 
%Note that the reference numbering continues from the main text to the SM.
% In the example below, Refs. 4-10 were cited only in the SM.     
\end{acknowledgments}

\appendix
\section{\label{app:pdoc} ESTIMATION OF THE OPTICAL CONTRAST DUE TO EDL-MODULATION}
To estimate the change in the scattering amplitude due to accumulation of ions in the electric double layer (EDL), we calculate the ratio of the polarizability of the thin screening layer to the polarizability of the nanoparticle in the limit of very large surface potential $V\gg\kbt/e \approx 25~$mV. 
Because the EDL is much thinner than the wavelength of the incident light, the details of the charge distribution inside the EDL have a negligible influence on the ELS intensity. 
We consider the screening ions to be uniformly distributed in a layer of thickness $\ls \ll a$ with $a$ the nanosphere radius. 
This is a proper approximation for large surface potentials because screening is mostly due to the compact layer.
The total number of excess mono-valent ions $N$ necessary for screening the nanosphere is given by $N=CV/e$ with $C$ the Stern-layer capacitance, $C = 4\pi \epsilon \epsilon_0 a^2/\ls$, and $e$ the elementary charge.
Using the definition of the Bjerrum length $\lb = \frac{e^2}{4 \pi \epsilon \epsilon_0 \kbt}$ , we can write down the expression for N in a more insightful form
\begin{equation}
N = \frac{a^2}{\lb\ls} \frac{Ve}{\kbt}\, .
\end{equation}

Assuming a deeply subwavelength nanoparticle, we can use the Rayleigh approximation to calculate the polarizability. In the Rayleigh independent scattering regime, the polarizability of the combined system of the nanosphere and the EDL is a volumetric sum of its constituents. 
To estimate the optical polarizability of the EDL, we use the empirical linear relation between refractive index and salt concentration $n_\mathrm{mix} = n_w + K x_s$, with $x_s$ the ratio between number density of salt ions and solvent molecules\cite{an_combined_2015} and $n_w$ the refractive index of water. Note that the value of $K$ is only reported for neutral salt solutions. From polarizability considerations for alkali-halide salts used in this work, the concentration of the halide ions is the dominant term for determining the refractive index of the salt-water mixture. We therefore can approximately use the same coefficient $K$ and use the excess anion density instead of the salt density in the refractive index change. Since we are interested in the relative changes of the polarizability, we can use any system of units. The Rayleigh polarizability of a scatterer in cgs-units is defined as
\begin{equation}\label{app:define}
\alpha = 3 V \frac{m^2-1}{m^2+2} \, ,
\end{equation}
with $V$ the volume and $m$ the refractive indices of the object divided by that of the surrounding medium. 

To estimate the polarizability of the EDL, we have to subtract the polarizability of the neutral solvent, and to linear order with $x_s$ we obtain
\begin{equation}\label{app:EDL} 
\alpha_\mathrm{EDL} = 4\pi a^2 \ls \frac{2 K x_s}{n_w} \, .
\end{equation}
We use the total number of excess ions at a given electrode potential to estimate the ratio 
\begin{equation}\label{app:xs} 
x_s = \frac{N}{4\pi a^2 \ls \rho_w} \, .
\end{equation}
where $\rho_w$ is the number density of water molecules and we have again used the assumption that screening charges are uniformly distributed in a layer of thickness $\ls \ll a$. By combining Eqs. (\ref{app:EDL}) and (\ref{app:xs}) we arrive at a simple formula for the EDL polarizability
\begin{equation}\label{app:EDLsimple}
\alpha_\mathrm{EDL} = \frac{2K N}{n_w \rho_w} \, .
\end{equation}
It is worth mentioning that the EDL polarizability depends only on the total number of excess ions inside the double layer in the Rayleigh approximation we have used. 
Finally, we can combine this polarizability to that of the spherical scatterer to obtain Eq.~(\ref{eq:aEDL}) of the main text.

\section{\label{app:saltstable} OPTICAL PROPERTIES OF ALKALI-HALIDE SALTS}
We could not find any measurement in the literature for the refractive index of solvated ions in the EDL. The anion polarizability of the halide anion~\cite{nagle_polarizabilities} and bulk refractive index of the solid salt~\cite{rii}, collected in the following table, can be used as rough guide for comparing the measured potentiodynamic contrasts for the three salts used in this report.

\begin{table}[htp]
\caption{Anion polarizability and solid refractive indices of Alkali-Halide salts}
\begin{center}
\begin{tabular}{|c|c|c|}
\hline
salt & anion polarizability & solid refractive index \\
\hline
NaCl & 14.6 & 1.54 \\
NaBr & 21 & 1.64 \\
NaI & 32 & 1.78 \\
\hline
\end{tabular}
\end{center}
\label{table1}
\end{table}%

%\nocite{*}

%\bibliographystyle{apsrev4-1}
\bibliography{PDSM_refs}

\end{document}